\documentclass[aps,prc,preprint,showpacs,floatfix,superscriptaddress]{revtex4-1}  
\usepackage{graphicx}
\usepackage{amsmath}

\newcommand {\nc} {\newcommand}
\nc {\beq} {\begin{eqnarray}} \nc {\eol} {\nonumber \\} \nc {\eeq}
{\end{eqnarray}} \nc {\eeqn} [1] {\label{#1} \end{eqnarray}} \nc
{\eoln} [1] {\label{#1} \\} \nc {\ve} [1] {\mbox{\boldmath $#1$}}
\nc {\rref} [1] {(\ref{#1})} \nc {\Eq} [1] {Eq.~(\ref{#1})} \nc
{\re} [1] {Ref.~\cite{#1}} \nc {\dem} {\mbox{$\frac{1}{2}$}} \nc
{\arrow} [2] {\mbox{$\mathop{\rightarrow}\limits_{#1 \rightarrow
#2}$}}
\begin{document}
\title{Reply to the "Comments to the paper "Detailed study of the astrophysical direct capture reaction $^{6}{\rm Li}(p, \gamma)^{7}{\rm
Be}$ in a potential model approach" " by S.B. Dubovichenko, A.S.
Tkachenko, R. Ya. Kezerashvili, arXiv:2401.04281 (2024)}

\author {E.M. Tursunov}
\email{tursune@inp.uz} \affiliation {Institute of Nuclear Physics,
Academy of Sciences, 100214, Ulugbek, Tashkent, Uzbekistan}
\affiliation {National University of Uzbekistan, 100174  Tashkent,
Uzbekistan}
\author {S.A. Turakulov}
\email{turakulov@inp.uz} \affiliation {Institute of Nuclear Physics,
Academy of Sciences, 100214, Ulugbek, Tashkent, Uzbekistan}
\author {K.I. Tursunmakhatov}
\email{tursunmahatovkahramon@gmail.com} \affiliation {Gulistan State
University,  120100, Gulistan, Uzbekistan}

\begin{abstract}
The differences in potential models used in our work and in the
original paper of the authors of the Comment are discussed. The
neglecting of simple rules of reaction calculations is shown as a
possible origin of the defect in the temperature dependence of the
reaction rates in the work of the authors of the Comment.

\end{abstract}


\maketitle

As the authors of the original paper "Detailed study of the
astrophysical direct capture reaction $^{6}$Li(p,$\gamma)^{7}$Be in
a potential model approach" published in  Ref.~\cite{tur2023} we
confirm that there are several points in the Comments which should
be taken into the consideration.

\par Firstly, the Gaussian potentials are widely used
by many authors in nuclear, atomic, and particle physics. In nuclear
physics, the most famous $\alpha-\alpha$ potential was published a
long time ago in Ref.~\cite{buck77}. The authors of the Comment
stated that the $p+^6\rm {Li}$ potentials in Ref.~\cite{tur2023} are
slightly modified versions of their original potentials from Ref.
~\cite{dub2022}. However, as can be found in Ref.~\cite{tur2023},
the parameters of the Gaussian potentials in the most important
initial $S$-wave and final $P$-wave states have been adjusted to the
experimental data, deduced from recent research collaboration
reports.

\par The $S$-wave Gaussian potential of Ref.~\cite{tur2023} with parameters $V_0$=-52.0 MeV and
$\alpha_0$=0.297 fm$^{-2}$ were adjusted to reproduce the low-energy
astrophysical S-factor of the LUNA collaboration published in Ref.
~\cite{LUNA2020}. This potential additionally reproduces the
experimental phase shifts at low energies of Ref.~\cite{skill95}. On
the other hand, the parameters of the $S$-wave potential of
Ref.~\cite{dub2022} are different: $V_0$=-58.0 MeV and
$\alpha_0$=0.4 fm$^{-2}$.   The most important point here is that
these potentials yield quite different $S$-wave scattering lengths,
$a_{01}$=10.01 fm of Ref.~\cite{tur2023} and $a_{01}$=11.45 fm of
Ref.~\cite{dub2022}. Since the $S$-wave scattering cross-section at
low energies is defined by the square of the scattering length, the
above potentials differ by about 30\% in the description of the
scattering cross-section. The parameters of the final state
$^2P_{3/2}$-wave ($V_0$= -76.6277 MeV, $\alpha_0$=0.175 fm$^{-2}$)
and $^2P_{1/2}$-wave ($V_0$= -74.8169 MeV, $\alpha_0$=0.1731
fm$^{-2}$) potentials of Ref.~\cite{tur2023} have been adjusted to
new central empirical squared ANC values of 4.81 fm$^{-1}$ and 4.29
fm$^{-1}$ from Ref.~\cite{kiss2021} for the $^7$Be(3/2$^-$) ground
and $^7$Be(1/2$^-$) excited bound states, respectively. The closest
versions (2 and 5) of the potential model of Ref.~\cite{dub2022}
yield the corresponding estimates of 5.1076 fm$^{-1}$ and 4.6656
fm$^{-1}$, respectively. Since the astrophysical S-factor is defined
by the square of the ANC, one can easily find that the differences
in the corresponding values of Ref.~\cite{tur2023} and
Ref.~\cite{dub2022} are of order 20\%.

\par Another important point is that the correct
description of the direct LUNA data for the astrophysical $S$-factor
and empirical reaction rates requires a reproduction of  both
absolute values and energy dependence of the $S$-factor and
temperature dependence of the reaction rates. A reproduction of the
temperature dependence means that the theoretical curve should be
parallel to the empirical curve. Although the theoretical curve of
Ref.~\cite{dub2022} lies close to the empirical reaction rates of
the LUNA collaboration inside the error bar, they are not parallel
each to other. A possible reason could be the neglecting of simple
rules of  reaction calculations in Ref.~\cite{dub2022}. Namely, the
authors use the atomic mass units, $m_p$=1.00727647 amu,
m($^6$Li)=6.01347746 amu for the calculations of the astrophysical
$S$-factor and the reaction rates. However, they put $\hbar^2/m_0$ =
41.4686 MeV fm$^2$  instead of $\hbar^2/m_0$ =41.8016 MeV fm$^2$,
which is not consistent with the chosen system of units. Our
calculations show that such a drawback seriously affects the final
results for the astrophysical S-factor and reaction rates.


\begin{thebibliography} {*}

\bibitem{tur2023} E.M. Tursunov, S.A. Turakulov, and K.I. Tursunmakhatov, {\it Phys. Rev. C}  {\bf 108}, 065801 (2023).
\bibitem{buck77} B. Buck, H. Friedrich, and C. Wheatley, {\it Nucl. Phys. A} {\bf 275},
246 (1977).
\bibitem{dub2022} S.B. Dubovichenko, A.S. Tkachenko, R.Y. Kezerashvili,
N.A. Burkova, and A.V. Dzhazairov-Kakhramanov, {\it  Phys. Rev. C}
{\bf 105}, 065806 (2022).
\bibitem{LUNA2020} LUNA Collaboration  (D. Piatti, T. Chillery, R. Depalo, M. Aliotta, D. Bemmerer, A. Best, A. Boeltzig, C. Broggini, C.G. Bruno, A. Caciolli
{\it et al}.),  {\it  Phys. Rev. C} {\bf 102}, 052802(R) (2020).
\bibitem{skill95} M. Skill, R. Baumann, G. Keil {\it et al}., {\it Nucl. Phys. A} {\bf 581}, 93
(1995).
\bibitem{kiss2021} G.G. Kiss, M. La Cognata, R. Yarmukhamedov, K.I.
Tursunmakhatov, I. Wiedenh\"over, L. T. Baby, S. Cherubini, A.
Cvetinovi\ifmmode \acute{c}\else \'{c}\fi{}, G. D'Agata, P. Figuera
{\it et al}., {\it  Phys. Rev. C} {\bf 104}, 015807 (2021).
\end{thebibliography}
 \end{document}